\documentclass[lettersize,journal]{IEEEtran}
\usepackage{amsmath,amsfonts}
\usepackage{algorithmic}
\usepackage{array}
\usepackage[caption=false,font=normalsize,labelfont=sf,textfont=sf]{subfig}
\usepackage{textcomp}
\usepackage{stfloats}
\usepackage{url}
\usepackage{verbatim}
\usepackage{graphicx}
\usepackage{bm}
\usepackage{cite}
\usepackage{booktabs}
\usepackage{tcolorbox} 
\usepackage{amssymb}
\usepackage{bbm}

\def\BibTeX{{\rm B\kern-.05em{\sc i\kern-.025em b}\kern-.08em
    T\kern-.1667em\lower.7ex\hbox{E}\kern-.125emX}}
\usepackage{balance}
\begin{document}
\bstctlcite{BSTcontrol}
\title{Beyond Task Relevance: Interaction-Aware Semantic Communication
for Wireless Agent Collaboration}

\author{Yiru Wang, Pengxin Guan, Xiumei Deng, Zehui Xiong,~\IEEEmembership{Senior~Member,~IEEE }

\thanks{Yiru Wang and Pengxin Guan are with the Department of Electronics, Peking University, Beijing, China (yiruwang.sg@gmail.com; guanpengxin@pku.edu.cn). Xiumei Deng is with the Information Systems Technology and Design Pillar, Singapore University of Technology and Design, Singapore (xiumei\_deng@sutd.edu.sg). Zehui Xiong is with the School of Electronics, Electrical Engineering and Computer Science (EEECS), Queen’s University Belfast, Belfast, BT7 1NN,
U.K. (z.xiong@qub.ac.uk). %\textit{(Corresponding author: Xiumei Deng)}
}}

\markboth{Journal of \LaTeX\ Class Files,~Vol.~18, No.~9, September~2020}%
{How to Use the IEEEtran \LaTeX \ Templates}

%TC:ignore
\maketitle

\begin{abstract}
Semantic communication in multi-agent systems often assumes that task-relevant information benefits the team once delivered. Yet semantic sharing can alter the receiver's decision and increase inter-agent dependence, potentially reducing team utility. Motivated by this phenomenon, we develop an interaction-aware framework for wireless collaboration between heterogeneous agents. We analytically relate communication-induced dependence to expected best-of-team utility and define a team-level semantic value for candidate messages. We consider semantic units at different abstraction levels and formulate a deadline- and power-constrained problem that maximizes predicted team-level net surplus. We derive an interaction-aware policy that selects at most one unit and its transmit power, or chooses not to transmit. Experiments on two verifier-scored optimization tasks show that full-solution exchange induces stronger structural convergence and more adverse team-level effects, whereas method hints exhibit pronounced directionality. Across the evaluated power budgets, our policy achieves the highest net team-level surplus while consuming less energy than semantic-aware and full-solution baselines.
\end{abstract}

%TC:endignore

\section{Introduction}
\IEEEPARstart{S}{emantic} communication (SemCom) has emerged as a
promising paradigm for next-generation wireless networks by shifting
the communication objective from accurate bit recovery toward the
delivery of task-relevant meaning. Early learning-based SemCom systems
demonstrated that semantic representations can improve transmission
efficiency and robustness over constrained wireless channels
\cite{xie2021deep}. This paradigm has subsequently evolved toward
task- and goal-oriented communication, where information is selected
and transmitted according to its contribution to downstream tasks
rather than source-level fidelity
\cite{xie2022task,zhou2024goal}. 

Meanwhile, advances in large artificial intelligence (AI) models are transforming wireless
terminals from passive data consumers into autonomous agents capable
of perception, reasoning, planning, and decision-making. Recent studies
have explored large-model-powered agents for intelligent
communications and AI-native edge systems, where agents interact with
wireless infrastructures and other agents to accomplish complex tasks
\cite{jiang2026agentic,li2025aaas}. In such systems, communication no
longer merely delivers semantic features to a downstream decoder.
Instead, the received information can directly influence an agent's
subsequent reasoning and action. %This distinction is particularly important for heterogeneous agents, whose independently generated solutions may provide complementary capabilities and solution diversity.

The effect of inter-agent information exchange, however, is not always
beneficial. Studies of multi-agent reasoning have shown that the value
of interaction depends strongly on how agents communicate: multi-agent
debate does not consistently outperform independent or ensemble
reasoning, while communication topology can affect both
performance and efficiency \cite{smit2024mad,li2024sparse}. More
recently, interaction tax was identified in large language model-based multi-agent reasoning, showing that exposing heterogeneous agents to one another's complete solutions can rapidly increase solution
similarity and erode the diversity advantage of independent reasoning
\cite{Interaction}.

This observation raises a question for wireless SemCom:
\emph{does information useful to a receiver necessarily improve the
team's best available solution?} Task relevance indicates whether a
message pertains to the receiver's task, while receiver-local gain
measures its effect on that agent. Neither alone determines its value
to a team. After reception, the
receiver's updated candidate may change both its quality and its
dependence on the sender's candidate, thereby changing expected
best-of-team utility. Existing SemCom designs often consider
task-related quality alongside bandwidth, latency, and energy
\cite{chaccour2025less,guo2025survey}, but these criteria alone do not
quantify this interaction-induced effect. Meanwhile, the interaction
tax in \cite{Interaction} identifies diversity loss at the reasoning
level without addressing semantic abstraction or wireless delivery.
We therefore ask how to value semantic units by their downstream
team-level effects and use that value to decide whether to transmit,
what to transmit, and at what power.

To address this question, we develop an interaction-aware semantic
communication framework for wireless heterogeneous-agent systems. Our contributions are as follows:
\begin{itemize}
    \item We establish an analytical framework that relates
    interaction-induced dependence to the expected best-of-team utility,
    and define an interaction-aware semantic value to quantify the
    downstream team-level effect of semantic exchange beyond
    receiver-local improvement.

    \item We characterize how semantic abstraction and communication
    direction affect multi-agent interactions. Experiments show that
    full-solution sharing induces the strongest convergence and the most
    adverse team-level effects, whereas method-level guidance exhibits
    pronounced directionality. 

    \item We formulate a deadline- and power-constrained problem that selects at most one semantic unit and its transmit power to maximize predicted net team-level surplus. Across the evaluated power budgets, the resulting policy outperforms semantic-aware and full-solution baselines in team-level surplus while consuming substantially less energy.
\end{itemize}

\section{System Model}
\subsection{Collaborative Multi-Agent System}

We consider a wireless multi-agent system consisting of a set of
heterogeneous agents $\mathcal{N}=\{1,\ldots,N\}$ that collaboratively
solve a common task. The agents may employ different foundation models,
reasoning strategies, or local observations, and therefore explore
different regions of the solution space. At the start of interaction round $t$, each agent $i\in\mathcal N$ holds a pre-transmission candidate solution $\mathbf{s}_i(t)$. Due to agent heterogeneity and stochastic generation, the candidate solutions produced by different agents can exhibit complementary characteristics, which provides the potential benefit of
multi-agent collaboration.

During task execution, agents can exchange task-related semantic information over wireless links. Instead of treating the complete candidate solution as the only transferable message, we represent the information available at agent $i$ as a set of semantic units $\mathcal{M}_i(t)=\{m_{i,1}(t),\ldots,m_{i,K_i}(t)\}$, where $K_i$ denotes the number of candidate semantic units. A semantic unit represents task information at a particular abstraction level, such as the transmitter's verifier-derived score or feedback, a method hint, a partial solution, or a complete solution. At each communication opportunity, the transmitting agent selects at most one semantic unit for transmission. Upon successful delivery, the receiver uses the received information, together with its existing task information and pre-interaction candidate, to generate an updated candidate solution. The receiver submits a single final candidate and cannot retrospectively select between its original and updated candidates based on their realized scores\footnote{The shared score or feedback pertains to the transmitter's candidate, whereas the receiver's candidate is evaluated only after the interaction stage.}.

Let $\mathcal{V}(\cdot)$ denote a task-specific evaluator for the candidate solution. The quality of the candidate maintained by agent $i$ at interaction round $t$ is
defined as
\begin{equation}
    S_i(t)
    =
    \mathcal{V}\!\left(\mathbf{s}_i(t)\right).
    \label{eq:candidate_quality}
\end{equation}

The evaluator can represent an objective function, constraint checker, or task reward, depending on the considered application. Due to model heterogeneity and stochastic generation, different agents may exhibit complementary solution-search behaviors and thus produce candidates with different quality distributions over interaction rounds.

Different from consensus-oriented multi-agent systems, the considered
system does not require all agents to converge to an identical solution.
Instead, the agents collaboratively seek at least one high-quality
candidate, and the best candidate can be selected according to the task
evaluator. The expected best-of-team utility \cite{Chow2025BoN} is defined as
\begin{equation}
    J(t)
    =
    \mathbb{E}
    \left[
        \max_{i\in\mathcal{N}} S_i(t)
    \right],
    \label{eq:team_utility}
\end{equation}
where the expectation is taken over the stochastic candidate-generation
and interaction processes at round $t$. This objective captures the solution-coverage
benefit of heterogeneous agents: even when no individual agent is
consistently superior, complementary search behaviors can increase the
expected quality of the best solution found by the team.

For analytical tractability, we focus on a representative pair of
agents $i$ and $j$. At round $t$, let
    $\mu_i(t)=\mathbb{E}[S_i(t)],$
    $\mu_j(t)=\mathbb{E}[S_j(t)],$
and let $\sigma_i^2(t)$ and $\sigma_j^2(t)$ denote the corresponding
variances. The associated pairwise collaborative utility is
$J_{ij}(t)
    =
    \mathbb{E}
    \left[
        \max\!\left(S_i(t),S_j(t)\right)
    \right].$ Their quality correlation is defined as
$\rho_{ij}(t)
    =
    \frac{
        \mathrm{Cov}\!\left(S_i(t),S_j(t)\right)
    }{
        \sigma_i(t)\sigma_j(t)
    }.$

%The correlation $\rho_{ij}(t)$ characterizes the statistical dependence between the solution qualities generated by the two agents. Semantic communication may improve the receiving agent's subsequent candidate, while also altering this dependence by steering its future search toward that of the transmitting agent. Therefore, the task-level effect of semantic sharing depends on both its individual quality improvement and its impact on collaborative exploration. 

\subsection{Wireless Link}

Consider the transmission of semantic unit $m_{i,k}$ from agent $i$
to agent $j$ over a wireless link. We assume a block-fading channel,
where the channel coefficient remains constant during the transmission
of one semantic unit and may vary across transmission intervals. Let
$h_{ij}(t)$ denote the channel coefficient at interaction round $t$ and
$p_{i,k}(t)$ denote the transmit power allocated to $m_{i,k}$. The
received signal-to-noise ratio (SNR) is given by
$\gamma_{ij,k}(t)
    =
    \frac{
        p_{i,k}(t)|h_{ij}(t)|^2
    }{
        N_0 W
    },$
where $W$ denotes the channel bandwidth and $N_0$ is the noise power
spectral density. Accordingly, the achievable transmission rate is
$R_{ij,k}(t)
    =
    W\log_2
    \left(
        1+\gamma_{ij,k}(t)
    \right).$
    
For a semantic unit with payload size $b_{i,k}$ bits, the corresponding
transmission latency is
$\tau_{ij,k}(t)
    =
    \frac{
        b_{i,k}
    }{
        R_{ij,k}(t)
    },$
and the transmission energy consumption is
$E_{ij,k}(t)
    =
    p_{i,k}(t)\tau_{ij,k}(t).$

%Importantly, successful delivery does not necessarily imply a positive contribution to the collaborative objective in \eqref{eq:team_utility}. Once decoded, different semantic units may exert different influences on the subsequent solution generated by the receiving agent. Therefore, the transmission decision should account for both the radio cost in \eqref{eq:energy} and the task-level effect induced by the received semantic content. The latter will be characterized through the semantic interaction model in the next section.

\section{Semantic Interaction Cost}

\subsection{Communication-Induced Correlation}

A successfully received semantic message can affect the subsequent
solution generated by the receiving agent in two distinct ways. First,
useful semantic content, such as a failure cause or a verified hint, may
improve the quality of the receiver's next candidate. Second, exposing
the receiver to another agent's reasoning or solution may steer its
subsequent search toward that of the transmitter, thereby altering the
statistical dependence between their future solution qualities. We
model these two effects separately.

Consider a semantic unit $m_{i,k}(t)$ generated by agent $i$ and
successfully decoded by agent $j$ at interaction round $t$. We focus on
the one-step effect of this interaction on the receiving agent, while
the transmitter's current candidate is treated as the reference during
this communication event. The resulting candidate-quality update is
modeled as
\begin{equation}
    S_j(t+1)
    =
    \left(1-\alpha_{ij,k}(t)\right)S_j(t)
    +
    \alpha_{ij,k}(t)S_i(t)
    +
    \delta_{ij,k}(t),
    \label{eq:semantic_update}
\end{equation}
where $\alpha_{ij,k}(t)\in[0,1]$ captures the extent to which the
receiver follows the transmitter's current solution, and $\delta_{ij,k}(t)\geq0$ captures the nonnegative residual semantic
benefit beyond transmitter-induced influence. The nonnegativity
constraint isolates interaction-induced loss from intrinsically
harmful semantic content. For notational simplicity, we subsequently use
$S_j^{+}\triangleq S_j(t+1)$, $S_j\triangleq S_j(t)$, and
$S_i\triangleq S_i(t)$ in the one-step analysis whenever no ambiguity
arises. The two terms in \eqref{eq:semantic_update} capture different effects
of semantic sharing. Specifically, $\alpha_{ij,k}(t)$ characterizes the
extent to which the receiver's subsequent candidate follows the
transmitter's current candidate, whereas $\delta_{ij,k}(t)$ represents
the message-specific innovation that improves the receiver beyond such
behavioral imitation. 

This model allows semantic units with similar task relevance to exert
different downstream effects. For example, a concise failure indication
may provide a useful quality increment while only weakly steering the receiver's search, whereas sharing a complete solution may induce a much
stronger coupling between the two agents. Hence, semantic value cannot
be characterized by individual relevance alone. %or quality improvement 

Taking the expectation of \eqref{eq:semantic_update} yields
$\mu_j(t+1)
    =
    \left(1-\alpha_{ij,k}(t)\right)\mu_j(t)
    +
    \alpha_{ij,k}(t)\mu_i(t)
    +
    \delta_{ij,k}(t).$
Accordingly, the individual quality gain induced by semantic unit
$m_{i,k}(t)$ is
\begin{equation}
    G_{ij,k}(t)
    \triangleq
    \mu_j(t+1)-\mu_j(t)
    =
    \alpha_{ij,k}(t)
    \left[\mu_i(t)-\mu_j(t)\right]
    +
    \delta_{ij,k}(t).
    \label{eq:individual_gain}
\end{equation}

Therefore, semantic sharing may increase the expected quality of the
receiving agent while simultaneously changing its statistical dependence
on the transmitting agent. 

In practice, $\alpha_{ij,k}(t)$ and $\delta_{ij,k}(t)$ are latent
interaction effects and cannot be separately inferred from a single
communication event. We therefore first estimate their group-level
effects from historical interactions with the same semantic-unit type
and comparable interaction contexts. Let $\mathcal{D}_g$ denote such
an interaction group. Their estimated effects are obtained as
\begin{equation}
\begin{aligned}
    (\bar{\alpha}_g,\bar{\delta}_g)
    =
    \arg\min_{\substack{0\leq\alpha\leq1\\ \delta\geq0}}
    \sum_{n\in\mathcal{D}_g}
    \left[
        \Delta S_j^{(n)}
        -
        \alpha\Delta S_{ij}^{(n)}
        -
        \delta
    \right]^2 ,
\end{aligned}
\label{eq:interaction_calibration}
\end{equation}
where
$\Delta S_j^{(n)}=S_j^{+,(n)}-S_j^{(n)}$ and
$\Delta S_{ij}^{(n)}=S_i^{(n)}-S_j^{(n)}$.
Here, $\bar{\alpha}_g$ captures transmitter-induced coupling and
$\bar{\delta}_g$ is the residual quality gain. For an interaction $(i,j,k,t)$ belonging to group $g$, we use $\hat{\alpha}_{ij,k}(t)=\bar{\alpha}_g$, $\hat{\delta}_{ij,k}(t)=\bar{\delta}_g$.

%including semantic type, agent scores, score gap, payload size, and historical pairwise statistics. The sigmoid and softplus mappings enforce the corresponding parameter constraints.

%\subsection{Communication-Induced Correlation}

%We next characterize how semantic sharing changes the statistical dependence between the transmitter and receiver. %We first analyze the structural effect using the latent interaction parameter $\alpha_{ij,k}$ and then provide its pre-transmission counterpart based on the estimated $\hat{\alpha}_{ij,k}$.

Let $\rho_{ij}^{+}
    \triangleq
    \mathrm{Corr}\!\left(S_i,S_j^{+}\right)$
denote the post-interaction correlation between the transmitter's
current candidate quality and the receiver's updated candidate quality. 
The post-interaction correlation is given by
\begin{equation}
    \rho_{ij}^{+}
=
\frac{
    (1-\alpha_{ij,k})\rho_{ij}\sigma_j
    +\alpha_{ij,k}\sigma_i
}{
    \sqrt{
        (1-\alpha_{ij,k})^{2}\sigma_j^{2}
        +\alpha_{ij,k}^{2}\sigma_i^{2}
        +2\alpha_{ij,k}(1-\alpha_{ij,k})
          \rho_{ij}\sigma_i\sigma_j
    }
}.
    \label{eq:post_corr}
\end{equation}

For $0\leq\rho_{ij}<1$ and $0\leq\alpha_{ij,k}\leq1$,
differentiating \eqref{eq:post_corr} gives
$\partial\rho_{ij}^{+}/\partial\alpha_{ij,k}\geq0$.
Hence, stronger semantic influence increases inter-agent statistical
dependence under the considered model. 
%Lightweight semantic guidance with a small $\alpha_{ij,k}$ preserves more of the receiver's independent search behavior, whereas highly influential semantic information can drive the two agents toward more strongly correlated outcomes.

For practical pre-transmission evaluation, the predicted
post-interaction correlation $ \hat{\rho}_{ij,k}^{+}$ can be obtained by replacing the latent
$\alpha_{ij,k}$ by its estimate
$\hat{\alpha}_{ij,k}$ in 
\eqref{eq:post_corr}. 

\subsection{Team-Level Interaction Cost}

% For analytical tractability, we approximate the candidate-quality pair $(S_i,S_j^+)$ as jointly Gaussian, which enables a closed-formcharacterization of the dependence--utility relationship.
Gaussian approximations have been used for statistical inference
on aggregate language-model evaluation scores~\cite{Miller2024ErrorBars}. For analytical tractability, we approximate the score pair $(S_i,S_j^+)$ as jointly Gaussian, which gives an analytical expression for the expected maximum score\footnote{Section \ref{sec:experiments} compares the resulting utility estimates with empirical values.}. Let their means be $\mu_X$ and $\mu_Y$, standard deviations be $\sigma_X$ and $\sigma_Y$,
and correlation coefficient be $\rho$. Define
$\theta(\rho)
    =
    \sqrt{
        \sigma_X^2+\sigma_Y^2
        -2\rho\sigma_X\sigma_Y
    }$,
    $d(\rho)
    =
    \frac{\mu_X-\mu_Y}{\theta(\rho)}.$
The expected best-of-two utility admits the standard closed-form
expression \cite{Clark1961}
\begin{equation}
\begin{aligned}
    \mathcal{J}
    (&\rho;\mu_X,\mu_Y,\sigma_X,\sigma_Y)
    \triangleq
    \mathbb{E}\!\left[\max(X,Y)\right] \\
    &=
    \mu_X\Phi\!\left(d(\rho)\right)
    +
    \mu_Y\Phi\!\left(-d(\rho)\right)
    +
    \theta(\rho)\phi\!\left(d(\rho)\right),
\end{aligned}
\label{eq:gaussian_team_utility}
\end{equation}
where $\Phi(\cdot)$ and $\phi(\cdot)$ denote the CDF and PDF of the
standard Gaussian distribution, respectively.

For fixed marginal distributions, differentiating
\eqref{eq:gaussian_team_utility} with respect to $\rho$ yields
$
    \frac{\partial \mathcal{J}}{\partial \rho}
    =
    -
    \frac{\sigma_X\sigma_Y}{\theta(\rho)}
    \phi\!\left(d(\rho)\right)
    <0,
$
for $\sigma_X,\sigma_Y>0$ and $-1<\rho<1$. Hence, stronger statistical
dependence reduces the expected best-of-team utility when the individual
quality distributions are fixed.

Let $\mu_j^{+}$ and $\sigma_j^{+}$ denote the post-interaction mean and
standard deviation of $S_j^{+}$. To isolate the loss caused by communication-induced dependence, define
the counterfactual utility
$J_{ij}^{\mathrm{cf}}
    \triangleq
    \mathcal{J}
    \left(
        \rho_{ij};
        \mu_i,\mu_j^{+},
        \sigma_i,\sigma_j^{+}
    \right),$
which preserves the post-interaction marginal qualities while
hypothetically retaining the pre-interaction correlation. In contrast,
the actual post-interaction utility is
$J_{ij}^{+}
    \triangleq
    \mathcal{J}
    \left(
        \rho_{ij}^{+};
        \mu_i,\mu_j^{+},
        \sigma_i,\sigma_j^{+}
    \right).$

We therefore define the team-level semantic interaction cost of semantic unit $m_{i,k}$ as
$C_{ij,k}^{\mathrm{int}}
    \triangleq
    J_{ij}^{\mathrm{cf}}
    -
    J_{ij}^{+}.$ Under the assumptions of  \eqref{eq:post_corr}, $0\leq\hat{\alpha}_{ij,k}(t)\leq1$ and hence $C_{ij,k}^{\rm int}\geq0$. Within the Gaussian model, this cost measures the utility difference due to the correlation change while holding the marginal score distributions after interaction fixed.

\section{Interaction-Aware Semantic Transmission}

\subsection{Net Semantic Value}

Communication can improve the receiver's marginal quality while
simultaneously reducing the value of independent exploration. We
therefore define the interaction-aware net semantic value of
$m_{i,k}$ as its net contribution to the collaborative
utility:
\begin{equation}
\begin{aligned}
    V_{ij,k}^{\mathrm{sem}}
    &\triangleq
    J_{ij}^{+}-J_{ij} =
    \left(J_{ij}^{\mathrm{cf}}-J_{ij}\right)
    -
    C_{ij,k}^{\mathrm{int}},
\end{aligned}
\label{eq:net_semantic_value}
\end{equation}
where $J_{ij}$ is the pre-interaction collaborative utility. Thus, $V_{ij,k}^{\rm sem}$ is the change in expected pairwise utility following successful reception, combining changes in the receiver's marginal score distribution and in score correlation. %A positive value indicates a utility gain before transmission costs are included.

Since the interaction effects are unknown before transmission, the
decision relies on the estimates $\hat{\alpha}_{ij,k}$ and
$\hat{\delta}_{ij,k}$ obtained from
\eqref{eq:interaction_calibration}. The predicted post-interaction
statistics are
\begin{equation}
\begin{aligned}
    \hat{\mu}_{j,k}^{+}
    &=
    (1-\hat{\alpha}_{ij,k})\mu_j
    +
    \hat{\alpha}_{ij,k}\mu_i
    +
    \hat{\delta}_{ij,k},
    \\
    (\hat{\sigma}_{j,k}^{+})^2
    &=
    (1-\hat{\alpha}_{ij,k})^2\sigma_j^2
    +
    \hat{\alpha}_{ij,k}^2\sigma_i^2 \\
    &\quad+
    2\hat{\alpha}_{ij,k}(1-\hat{\alpha}_{ij,k})
    \rho_{ij}\sigma_i\sigma_j .
\end{aligned}
\label{eq:predicted_marginal_statistics}
\end{equation}

Accordingly, the implementable post-interaction utility is
$\widehat{J}_{ij,k}^{+}
    =
    \mathcal{J}
    \left(
        \hat{\rho}_{ij,k}^{+};
        \mu_i,\hat{\mu}_{j,k}^{+},
        \sigma_i,\hat{\sigma}_{j,k}^{+}
    \right),$
and the implementable pre-transmission semantic value is
$\widehat{V}_{ij,k}^{\mathrm{sem}}
    \triangleq
    \widehat{J}_{ij,k}^{+}
    -
    J_{ij}.$
    
\subsection{Interaction-Aware Transmission Policy}

%At each communication opportunity, agent $i$ transmits at most one kind of semantic unit to agent $j$, or chooses not to transmit. Let $\mathcal{K}_{ij}(t)\subseteq\{1,\ldots,K_i\}$ index the available units, and let $k=0$ denote no transmission.  Define the predicted net surplus by $\widehat{\Psi}_{ij,k}(p,t) = \widehat{V}_{ij,k}^{\mathrm{sem}}(t) -\lambda p\tau_{ij,k}(p,t), k\in\mathcal{K}_{ij}(t),$ with $\widehat{\Psi}_{ij,0}(0,t)=0$ and $\lambda\geq0$. The interaction-aware transmission problem can be formulated as 

At each communication opportunity, agent $i$ decides whether to transmit one of its available semantic units to agent $j$ and, if so, at what power. Since team performance depends on the best current candidate across agents, the relevant task benefit is the predicted change in best-of-team utility. Transmission also consumes energy, so we evaluate this benefit net of its physical cost. Let $\mathcal{K}_{ij}(t)\subseteq{1,\ldots,K_i}$ index the available units. For $k\in\mathcal{K}_{ij}(t)$, we define the \textit{predicted team-level net surplus} as $\widehat{\Psi}_{ij,k}(p,t)=\widehat{V}_{ij,k}^{\mathrm{sem}}(t)-\lambda p\tau_{ij,k}(p,t)$, where $p\tau_{ij,k}(p,t)$ is the transmission energy and $\lambda\geq0$ weights energy against team utility. We use $k=0$ and $p=0$ to represent no transmission, with $\widehat{\Psi}_{ij,0}(0,t)=0$. The resulting problem jointly optimizes the unit index $k$ and transmit power $p$ under the deadline and power-budget constraints:
\begin{subequations}
\label{eq:semantic_transmission_problem}
\begin{align}
\max_{k,p}\quad&
\widehat{\Psi}_{ij,k}(p,t)
\label{eq:semantic_transmission_objective}\\
\mathrm{s.t.}\quad&
k\in\{0\}\cup\mathcal{K}_{ij}(t),
\label{eq:unit_constraint}\\
&p=0,\quad\text{if }k=0,
\label{eq:notx_constraint}\\
&0<p\leq P_{\max},\quad\text{if }k\in\mathcal{K}_{ij}(t),
\label{eq:power_constraint}\\
&\tau_{ij,k}(p,t)\leq T_{\max},
\quad\text{if }k\in\mathcal{K}_{ij}(t).
\label{eq:latency_constraint}
\end{align}
\end{subequations}

For each unit, the minimum power satisfying the deadline is
$p_{ij,k}^{\mathrm{req}}(t)
=
\frac{N_0W}{g_{ij}(t)}
\left(
2^{b_{i,k}(t)/(WT_{\max})}-1
\right).$
Since $p\tau_{ij,k}(p,t)$ increases with $p$ under the rate model above,
$p_{ij,k}^{\mathrm{req}}(t)$ minimizes transmission energy among feasible
powers. Let
$\mathcal{K}_{ij}^{\mathrm{fea}}(t)
=\{k\in\mathcal{K}_{ij}(t):
p_{ij,k}^{\mathrm{req}}(t)\leq P_{\max}\}$ and set
$p_{ij,0}^{\mathrm{req}}(t)=0$. The interaction-aware transmission policy can be expressed as
\begin{equation}
\begin{aligned}
k^\star
&\in\arg\max_{k\in\{0\}\cup\mathcal{K}_{ij}^{\mathrm{fea}}(t)}
\widehat{\Psi}_{ij,k}
\!\left(p_{ij,k}^{\mathrm{req}}(t),t\right),\\
p^\star&=p_{ij,k^\star}^{\mathrm{req}}(t).
\end{aligned}
\label{eq:structured_rule}
\end{equation}
where $k^\star$ denotes the selected semantic unit ($k^\star=0$ for no transmission), $p^\star$ is its transmit power, and $p_{ij,0}^{\mathrm{req}}(t)\triangleq0$. Ties at zero surplus are resolved in favor of $k^\star=0$.

%Thus, a unit is transmitted only if its predicted semantic value exceeds the energy cost of meeting the deadline. 

\section{Experimental Results}
\label{sec:experiments}

\subsection{Experimental Setup}
%Each agent first generates an independent candidate, which is frozen before interaction. 
We evaluate Knapsack-50 \cite{Martello1990Knapsack} and MaxCut
\cite{Goemans1995MaxCut} using GPT-4.1-mini \cite{OpenAI2025GPT41}
and Gemini-2.5-Flash \cite{Google2025Gemini25Flash} as heterogeneous
agents. At each communication opportunity, we consider three candidate semantic units: S1 (Outcome Feedback), which contains verifier-derived feedback; S2 (Method Hint), which conveys a high-level strategy without reconstructive details; and S3 (Full Solution), which shares the complete candidate and verifier score
\footnote{
Our approach uses predefined agent pairs, task types, and
message-construction rules for each semantic abstraction level, with interaction effects estimated
separately for each group. Changes in these settings may
require re-estimation. Joint optimization across multiple
agent pairs or semantic units would involve coupled,
history-dependent interactions under shared wireless resources.
We leave adaptive estimation and joint optimization for future work.
}.

The dataset contains 180 interaction records from 30 task instances.
We assign all records from each instance to the same split: 20
instances (120 records) for fitting and 10 instances (60 records)
for held-out evaluation. Within each task $\times$ semantic-type
$\times$ direction group $g$, the training records determine
$(\bar{\alpha}_g,\bar{\delta}_g)$ and the pre-interaction score
statistics. These quantities yield the group-level estimate
$\widehat V_g^{\rm sem}=\widehat J_g^+-\widehat J_g$ used for
transmission decisions on held-out instances. Receiver post-interaction
scores are not used in those decisions\footnote{To assess the Gaussian approximation of absolute best-of-team utility and the robustness of the dependence--utility relationship, we compare the closed-form and empirical utilities, obtaining a correlation of $0.977$ and a mean absolute difference of $1.8$ score points. Under permutations that preserve the empirical marginals while re-pairing sender and receiver-post scores, stronger dependence reduces empirical best-of-team utility in all $12$ groups.}.

%We evaluate current-round team utility using one active candidate per agent. After a successful interaction, the receiver's updated candidate replaces its original candidate for team selection, while the original candidate defines the pre-interaction baseline.
Candidate quality is normalized by the certified task optimum and
reported on a 0--100 scale. The realized team-value change is
$v_{\rm inst}\triangleq
\max(S_i,S_j^+)-\max(S_i,S_j)$.
We measure structural convergence by
$G_{\rm struct}\triangleq
D(\mathbf{s}_i,\mathbf{s}_j)-
D(\mathbf{s}_i,\mathbf{s}_j^+)$,
using Jaccard distance for Knapsack-50 and label-invariant normalized
Hamming distance for MaxCut.

We use a 2.4 GHz carrier, $W=100$ kHz bandwidth, a
$T_{\max}=10$ ms transmission deadline, and a thermal-noise power
spectral density of $-174$ dBm/Hz. Link distances range from 50 to
100 m. We model large-scale attenuation using log-distance path loss with
log-normal shadowing:
$PL(d)=PL_0+10\eta\log_{10}\!\left(\frac{d}{d_0}\right)+X_\sigma,
    X_\sigma\sim\mathcal{N}(0,\sigma_{\rm sh}^{2}).$
Here, $d$ is the transmitter--receiver distance. We set
$d_0=1$ m, $PL_0=40.05$ dB, $\eta=3.5$, and
$\sigma_{\rm sh}=6$ dB.
Small-scale fading is modeled as block Rayleigh fading. A selected message is treated as delivered when its transmission time meets the deadline.

\subsection{Performance Analysis}

\begin{figure}
	\centering
	\includegraphics[width=8cm]{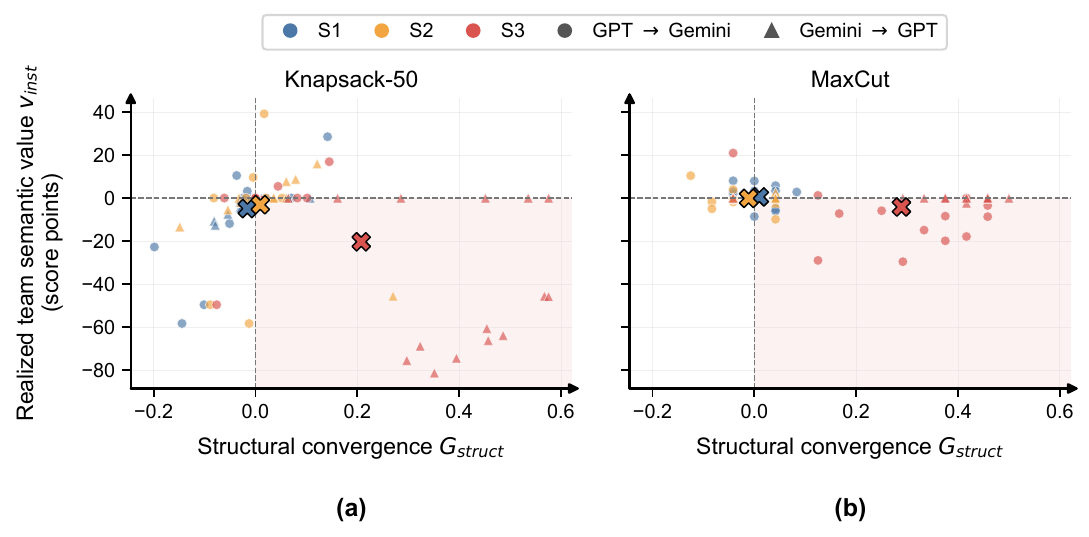}
	\caption{Joint landscape of the semantic interaction tax on (a) Knapsack-50 and (b) MaxCut. The shaded region indicates structural convergence with team-value degradation.}
	\label{fig:interaction_map}
\end{figure}

Fig.~\ref{fig:interaction_map} reveals the interaction-tax regime,
where structural convergence ($G_{\rm struct}>0$) coincides with
degraded team utility ($v_{\rm inst}<0$). S3 enters this regime much
more frequently than S1 and S2 and induces significantly stronger
convergence (Holm-adjusted $p<0.01$). The dominant harmful direction
also differs across tasks, highlighting the semantic- and
sender--receiver-dependent nature of the interaction tax.
%Therefore, full-solution sharing does not simply ``provide more useful information''; it can suppress the diversity of heterogeneous agents and pull a stronger receiver toward a less favorable solution. These observations motivate a receiver- and direction-aware semantic communication strategy rather than indiscriminate full-solution exchange.

\begin{figure}
	\centering
	\includegraphics[width=8cm]{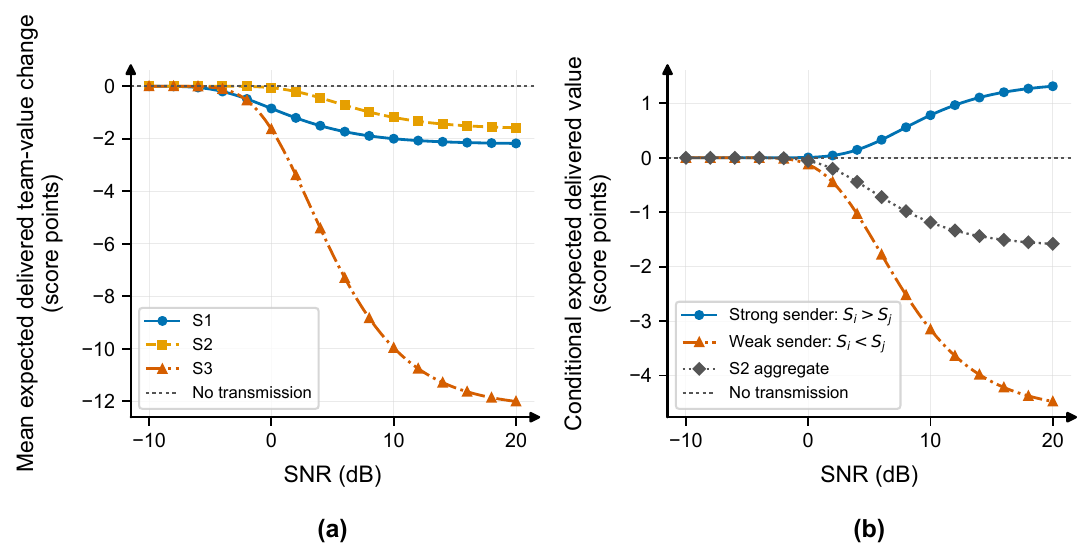}
	\caption{Wireless realization of semantic interaction effects: (a) delivered team-value change versus SNR for S1--S3; (b) strong-to-weak and weak-to-strong S2 interactions.}
	\label{fig:wireless_value}
\end{figure}

%Fig.~\ref{fig:wireless_value}(a) shows how the downstream interaction effect is progressively exposed as the wireless link becomes more reliable. At low SNR, most semantic packets fail to meet the transmission deadline. As the SNR increases, the three curves separate according to both their payload-dependent delivery reliability and their observed interaction values. In particular, S3 rapidly develops a large negative delivered value, indicating that a better channel can more reliably activate a harmful full-solution interaction. In contrast, S1 and S2 remain much closer to zero, with S2 exhibiting the most favorable aggregate behavior. This result highlights an important distinction: improving communication reliability does not necessarily improve team utility; rather, higher reliability amplifies the downstream effect, whether beneficial or detrimental, of the transmitted semantic information. In Fig.~\ref{fig:wireless_value}(b), we separate S2 interactions according to the initial sender--receiver quality ordering. When the sender initially outperforms the receiver, the delivered value increases with SNR and becomes approximately neutral-to-positive, whereas the opposite weak-to-strong direction becomes increasingly detrimental as reliable delivery takes effect. This contrast suggests that S2 can serve as a promising semantic abstraction for selective communication when guidance is delivered from a stronger agent to a weaker one.

Fig.~\ref{fig:wireless_value}(a) shows that higher reliability amplifies
the intrinsic effect of semantic interaction rather than necessarily
improving team utility. S3 becomes increasingly detrimental at high
SNR, whereas S1 and S2 remain much closer to neutral. Fig.~\ref{fig:wireless_value}(b) stratifies S2 interactions by pre-interaction score ordering. When $S_i>S_j$, expected delivered value is nonnegative because the sender's candidate remains available. A positive gain requires $S_j^+>S_i$. When $S_i<S_j$, expected delivered value becomes more negative with SNR as harmful updates are delivered more reliably. This motivates transmission decisions based on predicted team-level value.

\begin{figure}
	\centering
	\includegraphics[width=8cm]{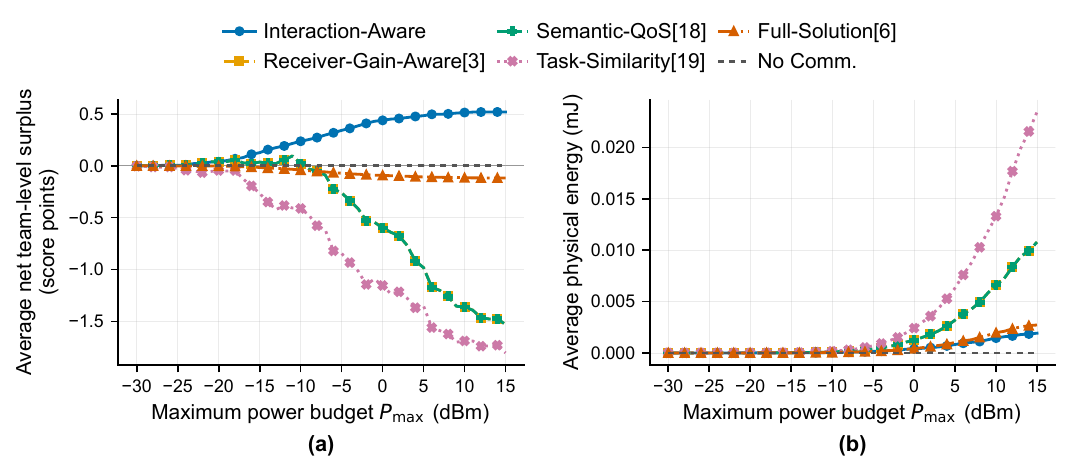}
	\caption{Average net team-level surplus and physical transmission energy of semantic transmission policies under varying maximum power budgets.}
	\label{fig:wireless_policy}
\end{figure}

In Fig.~\ref{fig:wireless_policy}, we compare Interaction-Aware with Receiver-Gain-Aware, which considers receiver-local task gain \cite{zhou2024goal}; Semantic-QoS-Aware, which minimizes energy subject to a receiver-local semantic-quality constraint \cite{xu2024semantic}; Task-Similarity-Aware, which trades task-description similarity against energy \cite{yan2022resource}; Full-Solution, which restricts communication to complete solutions \cite{smit2024mad}; and No Communication. All reported net team-level surplus values use a common energy price of $\lambda=672.74$ score points/J. Fig.~\ref{fig:wireless_policy}(a) shows that the mean net team-level surplus of Interaction-Aware rises from $0.0006$ at $-30$ dBm to $0.5187$ at $15$ dBm, where it approaches a plateau. As more transmissions become feasible, baselines that overlook communication-induced dependence increasingly select interactions that degrade team utility. In contrast, Interaction-Aware accounts for these downstream effects and avoids transmissions with negative predicted net value. Fig.~\ref{fig:wireless_policy}(b) shows that Interaction-Aware consumes $82.2\%$ and $91.8\%$ less energy than Semantic-QoS-Aware and Task-Similarity-Aware, respectively, at $15$ dBm, while retaining a positive mean net team-level surplus.

\section{Conclusion}
\label{sec:conclusion}

This paper develops an interaction-aware semantic communication framework by analytically characterizing how communication-induced dependence affects expected best-of-team utility. Building on this relationship, we formulate a deadline- and power-constrained problem to maximize predicted team-level utility gain net of weighted transmission energy. The resulting policy either remains silent or selects a semantic unit from different abstraction levels for transmission at an appropriate power. Experiments show that full-solution exchange induces the strongest structural convergence and more adverse team-level effects, whereas method hints exhibit direction-dependent effects while better preserving solution diversity. Across the evaluated power budgets, our policy achieves higher net team-level surplus with lower energy consumption than semantic-aware and full-solution baselines. 
%\section*{ACKNOWLEDGMENTS}

\bibliographystyle{IEEEtran}
\bibliography{IEEEabrv,myre}

\begin{thebibliography}{10}
\providecommand{\url}[1]{#1}
\csname url@samestyle\endcsname
\providecommand{\newblock}{\relax}
\providecommand{\bibinfo}[2]{#2}
\providecommand{\BIBentrySTDinterwordspacing}{\spaceskip=0pt\relax}
\providecommand{\BIBentryALTinterwordstretchfactor}{4}
\providecommand{\BIBentryALTinterwordspacing}{\spaceskip=\fontdimen2\font plus
\BIBentryALTinterwordstretchfactor\fontdimen3\font minus \fontdimen4\font\relax}
\providecommand{\BIBforeignlanguage}[2]{{%
\expandafter\ifx\csname l@#1\endcsname\relax
\typeout{** WARNING: IEEEtran.bst: No hyphenation pattern has been}%
\typeout{** loaded for the language `#1'. Using the pattern for}%
\typeout{** the default language instead.}%
\else
\language=\csname l@#1\endcsname
\fi
#2}}
\providecommand{\BIBdecl}{\relax}
\BIBdecl

\bibitem{xie2021deep}
H.~Xie, Z.~Qin, G.~Y. Li, and B.-H. Juang, ``Deep learning enabled semantic communication systems,'' \emph{IEEE Transactions on Signal Processing}, vol.~69, pp. 2663--2675, 2021.

\bibitem{xie2022task}
H.~Xie, Z.~Qin, X.~Tao, and K.~B. Letaief, ``Task-oriented multi-user semantic communications,'' \emph{IEEE Journal on Selected Areas in Communications}, vol.~40, no.~9, pp. 2584--2597, 2022.

\bibitem{zhou2024goal}
H.~Zhou, Y.~Deng, X.~Liu, N.~Pappas, and A.~Nallanathan, ``Goal-oriented semantic communications for {6G} networks,'' \emph{IEEE Internet of Things Magazine}, vol.~7, no.~5, pp. 104--110, 2024.

\bibitem{jiang2026agentic}
F.~Jiang, C.~Pan, K.~Wang, P.~Michiardi, O.~A. Dobre \emph{et~al.}, ``From large {AI} models to agentic {AI}: A tutorial on future intelligent communications,'' \emph{IEEE Journal on Selected Areas in Communications}, vol.~44, pp. 3507--3540, 2026.

\bibitem{li2025aaas}
B.~Li, T.~Liu, W.~Wang, C.~Zhao, and S.~Wang, ``Agent-as-a-service: An {AI}-native edge computing framework for {6G} networks,'' \emph{IEEE Network}, vol.~39, no.~2, pp. 44--51, 2025.

\bibitem{smit2024mad}
A.~P. Smit, N.~Grinsztajn, P.~Duckworth, T.~D. Barrett, and A.~Pretorius, ``Should we be going mad? a look at multi-agent debate strategies for {LLMs},'' in \emph{Proceedings of the 41st International Conference on Machine Learning}, 2024, pp. 45\,883--45\,905.

\bibitem{li2024sparse}
Y.~Li, Y.~Du, J.~Zhang, L.~Hou, P.~Grabowski \emph{et~al.}, ``Improving multi-agent debate with sparse communication topology,'' in \emph{Findings of the Association for Computational Linguistics: EMNLP 2024}, 2024, pp. 7281--7294.

\bibitem{Interaction}
S.~E. Ann, H.~Liu, and C.~Tan, ``The interaction tax: When communication erases diversity in multi-agent teams,'' in \emph{ICML 2026 Workshop: AI as a Tool for Mathematics, Computer Science, and Machine Learning}, 2026.

\bibitem{chaccour2025less}
C.~Chaccour, W.~Saad, M.~Debbah, Z.~Han, and H.~V. Poor, ``Less data, more knowledge: Building next-generation semantic communication networks,'' \emph{IEEE Communications Surveys \& Tutorials}, vol.~27, no.~1, pp. 37--76, 2025.

\bibitem{guo2025survey}
S.~Guo, Y.~Wang, N.~Zhang, Z.~Su, T.~H. Luan \emph{et~al.}, ``A survey on semantic communication networks: Architecture, security, and privacy,'' \emph{IEEE Communications Surveys \& Tutorials}, vol.~27, no.~5, pp. 2860--2894, 2025.

\bibitem{Chow2025BoN}
Y.~Chow, G.~Tennenholtz, I.~Gur, V.~Zhuang, B.~Dai \emph{et~al.}, ``Inference-aware fine-tuning for best-of-n sampling in large language models,'' in \emph{International Conference on Learning Representations}, 2025.

\bibitem{Miller2024ErrorBars}
E.~Miller, ``Adding error bars to evals: A statistical approach to language model evaluations,'' \emph{arXiv preprint arXiv:2411.00640}, 2024.

\bibitem{Clark1961}
C.~E. Clark, ``The greatest of a finite set of random variables,'' \emph{Operations Research}, vol.~9, no.~2, pp. 145--162, 1961.

\bibitem{Martello1990Knapsack}
S.~Martello and P.~Toth, \emph{Knapsack Problems: Algorithms and Computer Implementations}.\hskip 1em plus 0.5em minus 0.4em\relax Chichester, U.K.: John Wiley \& Sons, 1990.

\bibitem{Goemans1995MaxCut}
M.~X. Goemans and D.~P. Williamson, ``Improved approximation algorithms for maximum cut and satisfiability problems using semidefinite programming,'' \emph{Journal of the ACM}, vol.~42, no.~6, pp. 1115--1145, Nov. 1995.

\bibitem{OpenAI2025GPT41}
{OpenAI}, ``Introducing {GPT-4.1} in the {API},'' \url{https://openai.com/index/gpt-4-1/}, Apr. 2025, accessed: Sep. 10, 2026.

\bibitem{Google2025Gemini25Flash}
{Google DeepMind}, ``{Gemini 2.5 Flash},'' \url{https://ai.google.dev/gemini-api/docs/models/gemini-2.5-flash}, 2025, accessed: Sep. 10, 2026.

\bibitem{xu2024semantic}
C.~Xu, M.~B. Mashhadi, Y.~Ma, and R.~Tafazolli, ``Semantic-aware power allocation for generative semantic communications with foundation models,'' in \emph{2024 IEEE Globecom Workshops}, 2024, pp. 1--6.

\bibitem{yan2022resource}
L.~Yan, Z.~Qin, R.~Zhang, Y.~Li, and G.~Y. Li, ``Resource allocation for text semantic communications,'' \emph{IEEE Wireless Communications Letters}, vol.~11, no.~7, pp. 1394--1398, 2022.

\end{thebibliography}

\end{document}